\documentclass[10pt,prb,aps,twocolumn,showpacs,superscriptaddress]{revtex4}
\usepackage{graphicx}

\begin{document}

\title{Specific heat of quasi-2D antiferromagnetic Heisenberg models with 
varying inter-planar couplings}

\author{Pinaki Sengupta} 
\affiliation{Department of Physics, University of California,
Davis, California 95616}

\author{Anders W. Sandvik} 
\affiliation{Department of Physics, {\AA}bo Akademi University, 
Porthansgatan 3, FIN-20500, Turku, Finland}

\author{Rajiv R. P. Singh} 
\affiliation{Department of Physics, University of California,
Davis, California 95616}

\date{\today}

\begin{abstract}
We have used the stochastic series expansion (SSE) quantum Monte Carlo
(QMC) method to study the three-dimensional (3D) antiferromagnetic
 Heisenberg model on cubic lattices with in-plane coupling $J$ and
varying inter-plane coupling $J_\perp < J$. The specific heat curves
exhibit a 3D ordering peak as well as a broad maximum arising from short-range
2D order. For $J_\perp \ll J$, there is a clear separation of the two
peaks. In the simulations, the contributions to the total specific
heat from the ordering across and within the layers can be separated,
and this enables us to study in detail the 3D peak around $T_c$ (which
otherwise typically is dominated by statistical noise). We find that
the peak height decreases with decreasing $J_\perp$, becoming
nearly linear below $J_\perp = 0.2J$. The relevance of these results to
the lack of observed specific heat anomaly at the ordering transition
of some quasi-2D antiferromagnets is discussed.
\end{abstract}


\maketitle

\section{Introduction}

Spatially anisotropic systems and dimensional crossovers have been
issues of theoretical and experimental interest for many decades,
especially in context of classical critical phenomena.\cite{stanley1,stanley2} 
In recent years, a large number of quasi-low dimensional, low-spin, 
spatially anisotropic materials have
been synthesized and their properties investigated in great detail.
This has led to a renewed interest in these issues including the
role of enhanced quantum fluctuations.\cite{sudip,carlson,bocquet,affleck} 
Perhaps the most studied
of these are the cuprate family of materials, whose
parent stoichiometric compounds are antiferromagnetic insulators
which upon doping become high temperature superconductors. These are
layered compounds, where exchange coupling between the planes is many orders
of magnitude smaller than the exchange coupling in the 
planes.\cite{CHN,chubukov}
However, these are by no means the only systems where spatial anisotropy
and dimensional crossovers are important. The list of just 
novel transition-metal oxide materials, which despite their low-dimensionality
often develop 3-dimensional long-range order, includes
several cuprates, vanadates, copper-germenates, pnictide oxides, manganites,
etc.\cite{review}

In these materials both spatial anisotropy and anisotropy in spin-space
can be important in the development of 3D order. For example,
it is quite possible that in some cuprate families XY anisotropy
plays an important role in bringing about long-range order, while in
others it is the interplanar coupling which is primarily responsible
for the transition. Here, we will focus on layered systems with SU(2)
symmetry in spin-space. This is believed to be relevant to the material
La$_2$CuO$_4$. At the finite temperature 3D transition, one expects the
universality class for such a system to be that of classical 3D Heisenberg
model. However, in La$_2$CuO$_4$ no specific heat anomaly is seen at the 3D
transition,\cite{sun} contrary to expectations for the 3D classical Heisenberg model. In this paper we use a quantum Monte Carlo (QMC) method to verify 
that the transition in spatially anisotropic systems remains in the 
universality class of 3D classical Heisenberg model. Our primary goal is to 
clarify how the amplitude for the specific-heat anomaly at the transition 
is diminished in systems with weak interplanar couplings. This would help 
us predict which of the newly synthesized systems should show such anomalies,
given the finite experimental resolution.

A simple way to understand the reduction in the amplitude for the specific
heat anomaly, in these systems,
is to consider the effect of preexisting short-range order at the transition.
In a spatially anisotropic system, short range order in the planes can
develop at temperatures much above the 3D ordering temperature. And, if
the system is highly anisotropic, substantial spin-correlations can develop
in the planes before the eventual 3D transition. This means the effective
number of degrees of freedom involved in the 3D order is substantially
reduced. Hence, the specific-heat anomaly must diminish. Our goal is
to obtain a quantitative estimate for this effect.

The rest of the paper is organized as follows. We introduce the model and
the computational techniques used in Sec.~II. The results of the
simulations and the related discussions are presented in Secs.~III and
IV. We conclude in Sec.~V  with a summary of the results.

\section{Models and Simulation technique}

We have studied the Heisenberg antiferromagnet on an anisotropic cubic 
lattice. This model is given by the Hamiltonian
\begin{equation}
H=J\sum_{\langle i,j \rangle_{xy}}{\bf S}_i\cdot {\bf S}_j + J_\perp\sum_{\langle i,j \rangle_z}{\bf S}_i\cdot {\bf S}_j 
\end{equation} 
where $J$ ($J_\perp$) is the strength of the intra- (inter-) planar coupling.
The first (second) summation refers to summing over all nearest neighbors 
parallel (perpendicular) to the XY-plane. We will study the model as a
function of the dimensionless inter-plane coupling $\alpha=J_\perp/J$.

\begin{figure}
\includegraphics[width=8.3cm]{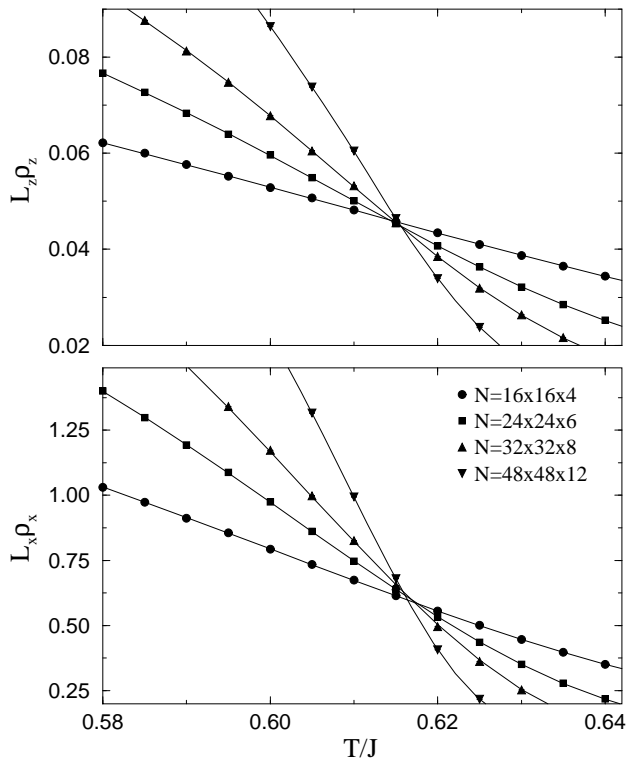}
\caption{Spin stiffness vs. temperature for different systems with the same
aspect ratio. The upper (lower) panel shows the stiffness perpendicular 
(parallel) to the planes.} 
\label{fig:rhos}
\end{figure}

The stochastic series expansion (SSE) method \cite{sse1,sse2} is a 
finite-temperature QMC technique based on importance sampling of the diagonal
matrix elements of the density matrix $e^{-\beta H}$. There are no 
approximations beyond statistical errors. Using the ``operator-loop'' 
cluster update,\cite{sse2} the autocorrelation time for the system sizes 
we consider here (up to $\approx 3\times 10^4$ spins) is at most a few Monte 
Carlo sweeps even at the critical temperature.\cite{dloops} 

On the dense temperature grids that we need in order to 
study the critical region in 
detail, we have further found that the statistics of the data obtained can
be significantly improved by the use of a tempering scheme.
\cite{tempering1,tempering2} A standard single-process tempering
method, where the temperature of the simulation fluctuates on a grid
of pre-selected temperatures, was previously used in a study of the 
isotropic 3D Heisenberg model.\cite{aws1} Here we use parallel tempering,
\cite{tempering2} where several simulations are run simultaneously on a 
parallel computer, using a fixed value of $\alpha$ and different, but closely
spaced, values of $T$ at and around the critical temperature. Along with the 
usual Monte Carlo updates, we attempt to swap the temperatures of SSE
configurations (processes) with adjacent values of $T$ at regular intervals 
(typically after every Monte Carlo step, each time attempting several 
hundred swaps) according to a scheme that maintains detailed balance in the 
space of the parallel simulations. This has favorable effects on the 
simulation dynamics, as the temperature of the SSE configurations will 
fluctuate across the critical temperature. More importantly in the case 
considered here, a given configuration will contribute to measured 
expectation values at several nearby temperatures, thereby reducing the
over-all statistical errors (at the cost of introducing correlations
between the errors, which is of minor significance here). Implementation 
of tempering schemes in the context of the SSE method have been discussed 
in Ref.~\onlinecite{bow}.

The thermodynamics of the 3D Heisenberg model on an isotropic simple cubic
lattice are fairly well understood from both analytic and computational
studies.\cite{baker,oitmaa,sauerwein} Recent large scale Monte Carlo 
studies\cite{aws1,dloops} have resulted in an accurate estimate of the 
critical temperature, $T_c/J \approx 0.946$. Several approximations also 
exist for $T_c$ of the anisotropic model.\cite{oguchi,liu,CHN,katanin} For 
weak coupling between the planes, the interplanar couplings can be treated 
in mean-field theory and lead to the relation 
$T_c \sim -1/\mbox{ln}(\alpha)$.\cite{CHN} We are not aware of any 
previous calculations of the specific heat of anisotropic systems.

\section{Locating the transition temperature}

We first determine the transition temperature for the model as a function 
of $\alpha$. An efficient way to do this is by studying the scaling properties
of the spin-stiffness. We have evaluated the spin stiffnesses both parallel 
to and perpendicular to the planes. The stiffness can be defined
\cite{kohn,kopietz} as the second derivative of the free energy with 
respect to a uniform
twist $\phi$: 
\begin{equation}
\rho = \frac{\partial ^2F(\phi)}{\partial\phi ^2}. 
\label{eq:rhoc}
\end{equation} 
The stiffness can also be related to the fluctuations of the ``winding number''
in the simulations \cite{pollock,harada,sse1,cuccoli} and hence can be 
estimated directly without actually including a twist. Since the twist can 
be applied parallel to or perpendicular to the planes, there are two 
different spin stiffnesses, $\rho_x$ and $\rho_z$, in the anisotropic
system considered here.

For a system of weakly coupled Heisenberg chains, it has been shown that
estimates for various observables for a spatially anisotropic system can
depend non-monotonically on the system size for square lattices.
\cite{multichain} One can instead use rectangular lattices to more rapidly
obtain monotonic behavior of the numerical results for extrapolating to
the thermodynamic limit. We expect similar effects in the present model 
at $\alpha \ll 1$. Hence we have studied tetragonal lattices with 
$L_x = L_y \neq L_z$. Lattices with an aspect ratio $R=L_x/L_z =4$ have 
been used to obtain the results presented here. We have chosen six different
values of $\alpha$, of the form of $\alpha=2^{-n}, n=1,\ldots,6$.

Following Ref.~\onlinecite{aws1}, we use the finite-size and temperature
dependence of the spin stiffnesses to determine the critical temperature.
\cite{barber} For a fixed aspect ratio, the stiffness at $T_{\rm c}$
is predicted to scale as
\begin{equation}
\rho_\mu = L_\mu^{2-d},~~~ \mu=x,z
\end{equation}
where $d$ is the dimensionality of the system. The above relation 
implies that for the 3D Heisenberg model, on a plot of $L_\mu\rho_\mu$ 
as a function of $T$ the curves for different system sizes will cross each
other at $T_c$. Results for $\alpha=1/4$ are shown in Fig.~\ref{fig:rhos}. 
The upper (lower) panel shows  $L_x\rho_x$ ($L_z\rho_z$) versus 
$T$ for four different system sizes. The curves indeed intersect each other 
almost at a single point. Subleading corrections are seen in the fact that 
the crossing points move slightly as the system size is increased. 
Interestingly, the behavior is opposite for the two stiffness constants;
in the case of $\rho_x$ the crossings move down in temperatures,
whereas the $\rho_z$ crossings move up. Hence, we believe that the
crossings for the two largest system sizes bracket the true $T_c$ and
we view them as the upper and lower bounds. From these results we 
estimate $T_c=0.6160 \pm 0.0005 $ for $\alpha=1/4$.

\begin{figure}
\includegraphics[width=8.3cm]{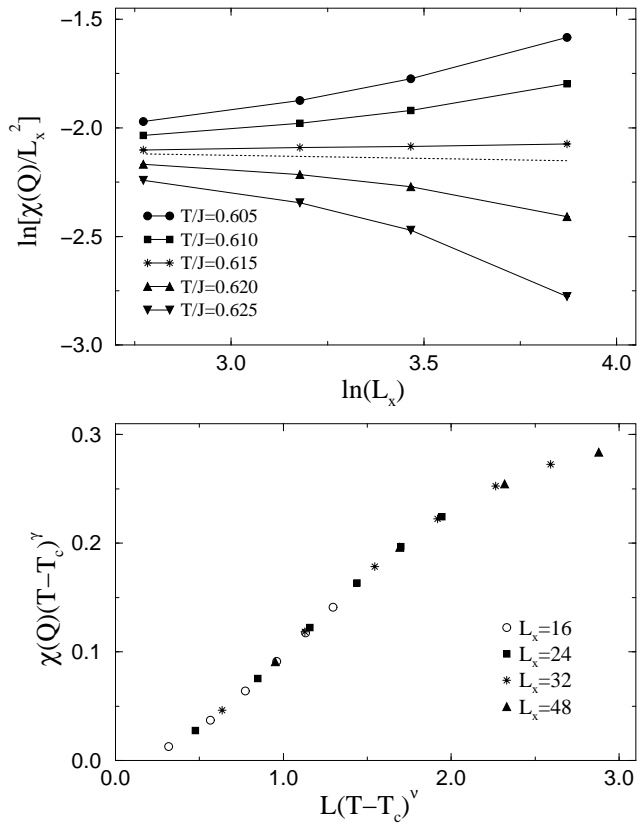}
\caption{Finite-size scaling of the staggered susceptibility at $\alpha=1/4$.
(a) Size dependence close to $T_c$. At $T_c$, the data are expected to fall 
on a straight line with slope $-\eta=-0.037$, which is indicated with the 
dotted line. (b) Scaling plot above $T_c$, using $T_c/J=0.616$ and the 3D 
classical Heisenberg exponents $\eta=0.037$ and $\nu=0.711$.} 
\label{fig:fss}
\end{figure}

Next we study the universality class of the transition. To this end, we 
consider the static magnetic susceptibility, defined as
\begin{equation}
\chi({\bf q})={1\over N}\sum_{\langle i,j \rangle}e^{i{\bf q}\cdot ({\bf r}_j-{\bf r}_j)}\int^{\beta}_0d\tau\langle S^z_j(\tau)S^z_i(0) \rangle,
\end{equation}
where $N=L_x^2L_y$ is the size of the system. At the critical 
temperature, the staggered susceptibility $\chi({\bf Q})$ should scale 
\cite{barber} with the system length as $L_x^{2-\eta}$, 
where ${\bf Q}=(\pi,\pi,\pi)$ is the 3D 
ordering wave vector. For any non-zero value of $J_{\perp}$, the transition 
is expected to belong to the classical 3D Heisenberg universality class, for 
which the critical exponents are known to a high degree of accuracy.
\cite{campostrini} The spin-spin correlation function exponent 
$\eta \approx 0.037$. Figure \ref{fig:fss}(a) shows $\alpha=1/4$ 
results for ${\mbox{ln}}(\chi({\bf Q})/L_x^2)$ 
versus ln($L_x$) at temperatures close to $T_c$. Asymptotically, we expect
the data to fall on a straight line with slope $-\eta \approx -0.037$ at 
$T=T_c$ and diverge upward (downward) for $T < T_c$ ($T > T_c$). This is 
indeed what we observe. The curves are completely consistent with the known 
value of $\eta$ and the estimate of $T_c$ obtained from Fig.~\ref{fig:rhos}.

We have also tested the expected scaling for $T > T_c$. In the thermodynamic 
limit, $\chi({\bf Q})$ should diverge as $t^{-\gamma}$, where $t=|T-T_c|$ 
and $\gamma=\nu(2-\eta)$. For a finite system, finite-size scaling predicts 
$\chi_L(t)=\chi_{\infty}(t)f[\xi(t)/L]$, with the correlation length 
diverging as $\xi\sim t^{-\nu}$. Hence on a plot of $\chi_L(t)t^{\gamma}$
versus $Lt^{\nu}$, data for different $L$ should collapse onto a single
curve. As shown in Fig.~\ref{fig:fss}(b), this is indeed the case with our 
estimated $T_c$ and the known 3D Heisenberg exponents.

\begin{figure}
\includegraphics[width=8.3cm]{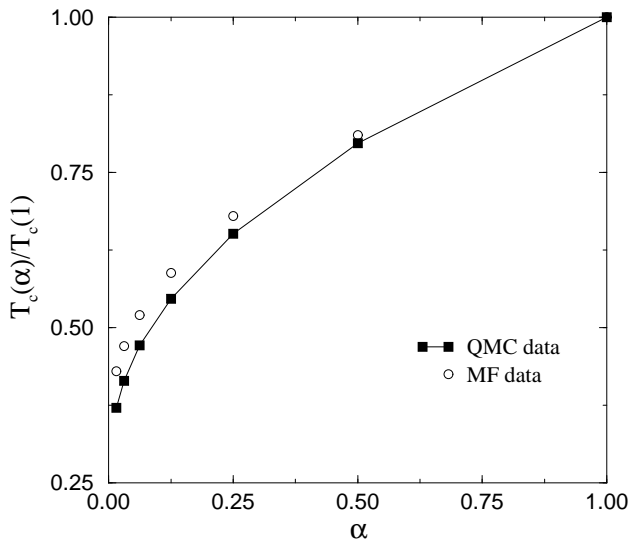}
\caption{Ratio of the critical temperature for the anisotropic system to that
for the isotropic system as a function of the anisotropy. The circles denote
the results from Eq.~\ref{eq:tc}.}
\label{fig:tc}
\end{figure}

We have here discussed the determination of $T_c$ and checked the consistency
with the expected universality class for $\alpha=1/4$. Using the spin 
stiffness scaling, we have located $T_c$ for several couplings $\alpha$.
The results are graphed in Fig.~\ref{fig:tc}. We compare our results
with the expression obtained by Liu,\cite{liu}
\begin{equation}
{1\over T_c}={1\over \pi^3}\int_0^\pi\int_0^\pi\int_0^\pi {\frac {dk_xdk_ydk_z}{2-\mbox{cos}k_x\mbox{cos}k_y+\alpha(1-\mbox{cos}k_z)}}.
\label{eq:tc}
\end{equation} 
We find that while this equation gives a
reasonable estimate for $T_c(\alpha)/T_c(1)$ for $\alpha$ close to unity, 
it begins to deviate substantially from the SSE results for small $\alpha$.

\section{Calculations of the specific heat}

Having determined $T_c$ as a function of $\alpha$, we now present the results
for the specific heat calculations. The specific heat is defined as the 
temperature derivative of the energy, $C_v = (\partial E/\partial T)/N$. 
As discussed in Appendix A, the SSE method allows us to obtain a direct 
estimate of the specific heat from the operator sequence in the simulation, 
so that any additional noise in the data due to numerical differentiation 
of the energy function can be avoided (although the two approaches in practice
give very similar results). The SSE estimator for the total specific heat
(i.e., not normalized by the lattice size) is
\begin{equation}
NC_v=\langle n^2\rangle - \langle n\rangle^2 - \langle n\rangle,
\label{cnn}
\end{equation}
where $n$ is the power-series expansion order (the number of bond operators
in the SSE operator string), which fluctuates in the
simulations. We will be interested in the contributions to $C_v$ from
the spin-spin ordering across and within the layers close to $T_c$.
Decomposing the Hamiltonian into an in-plane term $H_{p}$ and an
inter-layer term $H_z$, the specific heat
\begin{equation}
C_v = (\partial \langle H_{\rm p}\rangle/\partial T + 
\partial \langle H_z\rangle /\partial T)/N = C_v^{p}+C_v^z.
\label{cdecomp1}
\end{equation}
The SSE estimators for the two terms are given in terms of the numbers 
of bond operators in the expansion acting within a single layer ($n_p$)
and between two layers ($n_z$):
\begin{eqnarray}
NC^{p}_v & = &\langle n_p^2\rangle + \langle n_pn_z\rangle
 - \langle n_p\rangle^2 - \langle n_{p}\rangle \langle n_{z}\rangle
- \langle n_{p}\rangle, \label{cp} \\
NC^{z}_v & = &\langle n_{z}^2\rangle + \langle n_{p}n_z\rangle
 - \langle n_{z}\rangle^2 - \langle n_{p}\rangle \langle n_{z}\rangle 
- \langle n_{z}\rangle . \label{cz}
\end{eqnarray}
These expressions suggest the possibility of a different decomposition of
the specific heat. We will define $C_v^{\rm plane}$ as the part
of the estimator (\ref{cp}) that contains only purely in-plane 
contributions:
\begin{equation}
C^{\rm plane}_v =(\langle n_p^2\rangle 
 - \langle n_p\rangle^2 - \langle n_{p}\rangle )/N. \label{cplane}
\end{equation}
We refer to the remaining part of the total susceptibility as the 3D
contribution, i.e.,
\begin{equation}
C^{\rm 3D}_v = C_v - C^{\rm plane}_v = C^{\rm inter}_v + C^{\rm cross}_v 
\label{c3d},
\end{equation}
where the purely inter-plane contribution $C^{\rm inter}_v$ and cross-term
$C^{\rm cross}_v$ are given by
\begin{eqnarray}
C^{\rm inter}_v & = & (\langle n_z^2\rangle - \langle n_z\rangle^2 - 
\langle n_{z}\rangle)/N , \label{cinter} \\
C^{\rm cross}_v & = & 2(\langle n_{p}n_z\rangle
 - \langle n_{p}\rangle \langle n_{z}\rangle )/N , \label{cross}
\end{eqnarray}
We will show that the cross-term, half of which appears both in Eq.~(\ref{cp}) 
and Eq.~(\ref{cz}), dominates in the 3D contribution (\ref{c3d}). The
advantage of considering separately the different contributions to
$C_v$, either in the form of (\ref{cdecomp1}) or (\ref{c3d}) and, 
is that the full specific heat is dominated by the in-plane 
term and the other contributions can be difficult to discern due to 
statistical fluctuations. We will here focus in particular on the 3D 
contribution (\ref{c3d}).

\begin{figure}
\includegraphics[width=8.3cm]{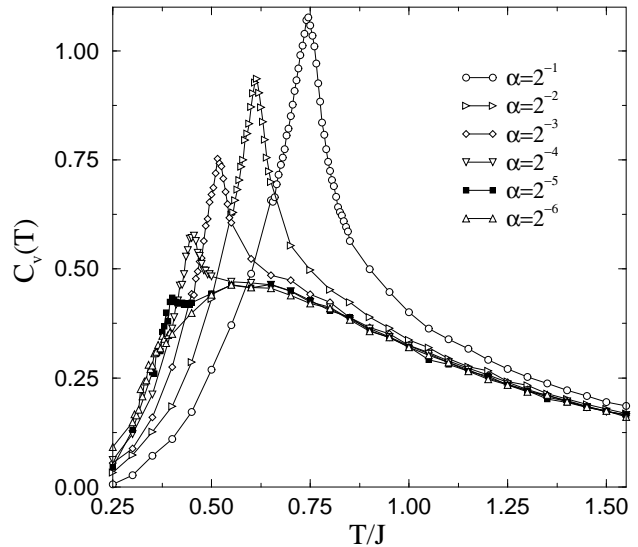}
\caption{The specific heat over a wide range of temperature for several
different anisotropies. The system size is $48\times 48\times 12$. The 
separation of the 3D ordering peak from the broad maximum arising out of 
the 2D physics is clearly visible for $\alpha \leq 2^{-3}$.} 
\label{fig:cvt}
\end{figure}

The specific heat for the 3D Heisenberg model on highly anisotropic lattices
($\alpha \ll 1$) will have two separate peaks, reflecting the 2D physics and 
the 3D ordering. The Mermin-Wagner theorem dictates that there can be no 
long-range order at $T>0$ in a strictly 2D system with a continuous symmetry.
The correlation length then diverges exponentially \cite{CHN} as $T\to 0$, 
and the specific heat has a broad maximum at $T/J\approx 0.7$.\cite{c2dcalc} 
This broad maximum is the dominant feature of the specific heat curve also for
small inter-planar couplings. On the other hand, for any $\alpha > 0$ there
is a phase transition to an ordered state at $T_c > 0$, as we have discussed 
in Sec.~III. The signature of this phase transition in the specific heat 
should be a peak at $T_c$. Since the transition belongs to the 3D Heisenberg 
universality class, there should a cusp-like singularity 
(instead of a divergent singularity) and the peak height is finite. 

SSE results for the specific heat over a wide temperature range are 
shown in Fig.~\ref{fig:cvt} for a system of size $N=48\times 48\times 12$. 
The effects of finite system size on the position of the peak and the peak 
height will be discussed later. The separation of the 3D ordering peak from
the broad maximum arising out of the 2D physics is clearly seen for 
$\alpha \leq 2^{-3}$. It is also seen that the excess peak height over the 
2D background decreases rapidly with decreasing $\alpha$, becoming hard to 
discern for $\alpha < 2^{-5}$.

\begin{figure}
\includegraphics[width=8.3cm]{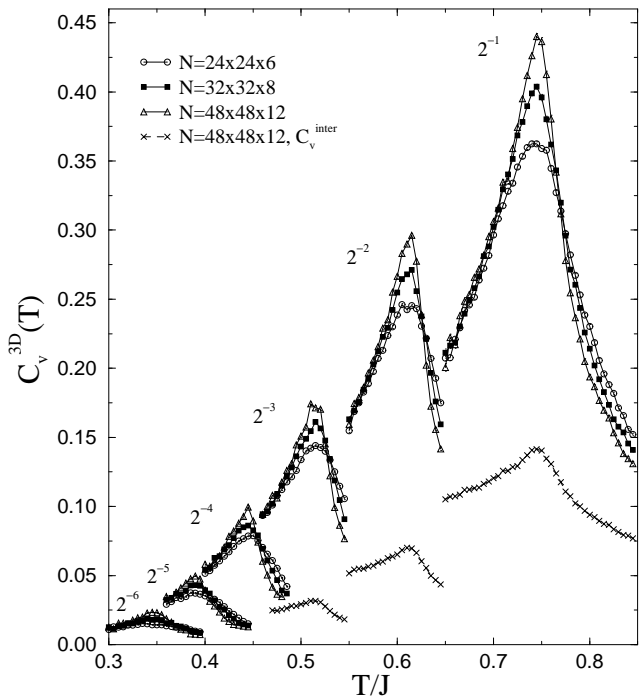}
\caption{The 3D contribution to the specific heat for several different
anisotropies, $\alpha$, and for three different system sizes. Results for
the purely inter-plane term $C_v^{\rm inter}$ are also shown for the three 
largest couplings (for the largest system size only).} 
\label{fig:cvx}
\end{figure}

Since the specific heat curve is dominated by its 2D contribution
when $\alpha \ll 1$, it is extremely difficult to study the nature of the 
3D peak near $T_c$. However, the 3D contribution (\ref{c3d}) can be studied 
to a high degree of accuracy. Results for several couplings $\alpha$ and
system sizes are shown in Fig.~\ref{fig:cvx}. Several features are immediately
apparent. The 3D contribution peaks at the N{\'e}el temperature and rapidly 
decreases away from it. The peak position moves only slightly with increasing
system size. The estimates of $T_c$ obtained from the position of the peaks 
are in close agreement with the more accurate estimates we obtained in Sec.~III
using the spin stiffness. In Fig.~\ref{fig:cvx} we also show some results
for the purely inter-plane contribution $C_v^{\rm inter}$ to $C_v^{\rm 3D}$, 
which is seen to be small and decreasing relative to the full 3D contribution
as $\alpha \to 0$. This is expected, as the estimator (\ref{cinter}) 
implicitly contains a prefactor proportional to $\alpha^2$, whereas the
cross-term (\ref{cross}) contains a linear $\alpha$ dependence. 

\begin{figure}
\includegraphics[width=8.3cm]{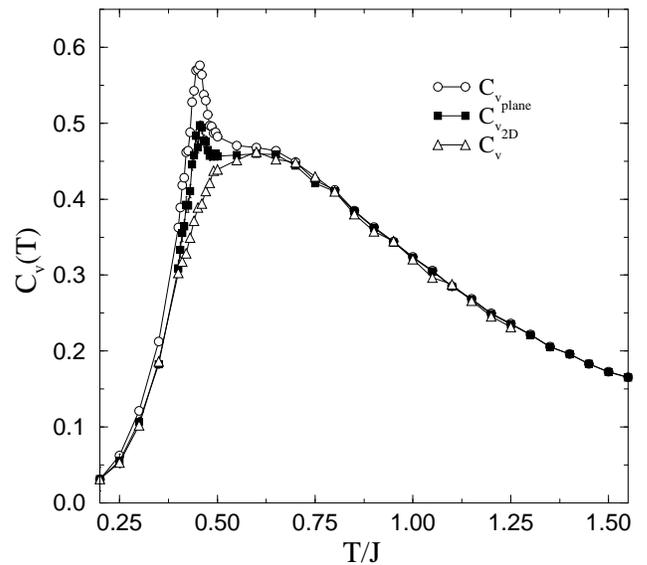}
\caption{The specific heat and its in-plane contribution for 
$\alpha=2^{-4}$. The anomalies at the transition temperature is clearly 
visible for both. The system size is $48\times 48\times 12$. For comparison, 
the specific heat for the pure 2D Heisenberg model is also shown.} 
\label{fig:cv2d}
\end{figure}

\begin{figure}
\includegraphics[width=8.3cm]{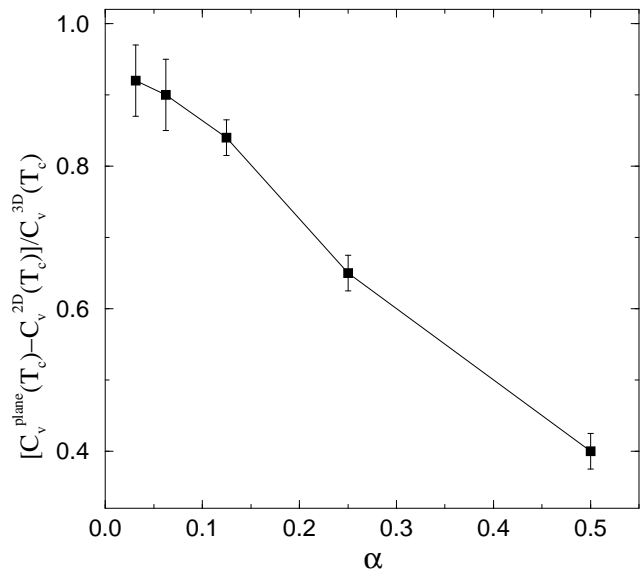}
\caption{The excess in $C_v^{\mbox{\tiny{plane}}}(T_c)$ over the 
specific heat of the pure 2D Heisenberg model at the 3D $T_c$, 
normalized by the corresponding 3D contribution to the total specific 
heat. The system size is $48\times 48\times 12$.} 
\label{fig:cv2drat}
\end{figure}

While the specific heat anomaly is most pronounced in the 3D contribution, it
is also present in the purely in-plane term. This is shown in 
Fig.~\ref{fig:cv2d}, where we have graphed the total specific heat and the 
purely in-plane contribution at $\alpha=1/16$, where the 3D ordering 
peak is well separated from the broad 2D maximum. We compare these results
with the specific heat $C_v^{\mbox{\tiny{2D}}}$ for a 2D system ($\alpha=0$).
As expected, the in-plane term for the 3D system is dominated by a broad 
maximum and coincides closely with the specific heat of the 2D system away 
from $T_c$. However, there is also a distinct peak at the 3D transition 
temperature. In order to quantify the relative sizes of the ordering peaks in
$C_v^{\rm 3D}$ and $C_v^{\rm plane}$, we next consider the excess at $T_c$
of the in-plane contribution over the specific heat of the pure 2D 
system model at the same temperature. Its ratio to the 3D contribution
is graphed as a function of the coupling $\alpha$ in Fig.~\ref{fig:cv2drat}. 
As $\alpha \to 0$, this ratio appears to converge to a value $\approx 1$, 
or, in other word, the ordering peak in the in-plane contribution
becomes nearly equal to that of the 3D contribution.

\begin{figure}
\includegraphics[width=8.3cm]{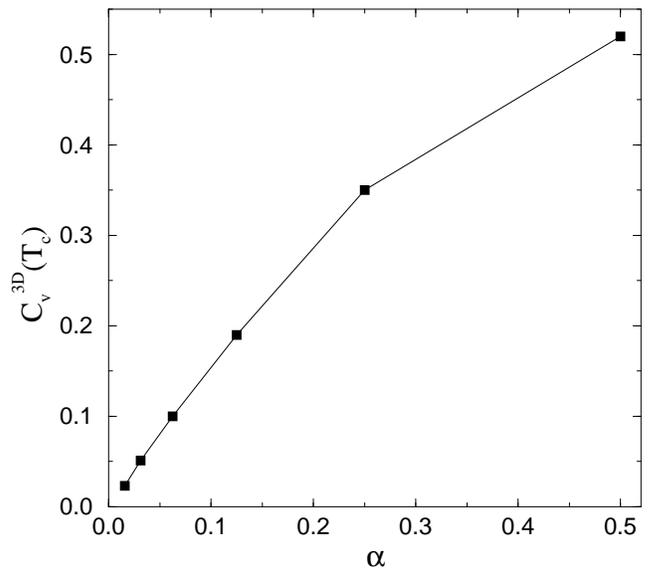}
\caption{The peak height for the 3D ordering extrapolated to the thermodynamic 
limit as a function of the anisotropy. For small anisotropies, the peak height
increases approximately linearly with the anisotropy.} 
\label{fig:pkht}
\end{figure}

The peak height $C_v^{\mbox{\tiny{3D}}}(T_c)$ decreases rapidly with 
decreasing $\alpha$. To get a more quantitative estimate of the nature of its
variation with $\alpha$, we have extracted the thermodynamic peak height 
for different $\alpha$. The specific heat exponent, which governs the
scaling of the peak to infinite size, is small (and negative),
\cite{campostrini} and the 
statistical errors of our data are relatively large for small $\alpha$.
The extrapolation is therefore affected by some uncertainty that is not
easy to quantify precisely. Our results are shown in Fig.~\ref{fig:pkht}. 
For small $\alpha$, the peak height is nearly linear in $\alpha$. This 
behavior can be roughly understood by the argument that the specific heat 
anomaly should scale as $1/\xi^2$, where $\xi$ is the correlation length of 
the $2D$ system at the 3D transition temperature. Furthermore, the 3D 
correlations become significant and lead to the 3D transition \cite{CHN} 
when $\xi^2\alpha\approx 1$. Thus the amplitude for the 
specific heat anomaly should vanish linearly with $\alpha$. It would be 
interesting to compare the specific heat anomaly of various quasi-2D 
Heisenberg systems against this result.

\section{Conclusions}

In this paper we have studied the 3D ordering transition in a model
of weakly coupled Heisenberg planes. Our results on the transition
temperature and universality class of the transition are in accord
with general expectations. Our primary focus here was on the specific
heat and in particular on the specific heat anomaly at the 3D
ordering transition. We find that for small $J_\perp$ the amplitude
for the specific heat anomaly is a nearly linear function of $J_\perp$.
It should be possible to compare this result directly
against experiments on various anisotropic materials.
However, it is clear that for highly anisotropic systems (such as
La$_2$CuO$_4$, where the anisotropy maybe as small as $10^{-6}$)
such anomalies will be very difficult to detect above the background.

\acknowledgments{
This work was supported in part by NSF grant number
DMR-9986948 (PS and RRPS) and by The Academy of Finland, project 
No.~26175 (AWS). Part of the simulations were carried out on the 
IBM SP machine at NERSC. 

\appendix
\section{The SSE method}

The SSE method has been discussed in several papers.\cite{sse1,sse2,dloops}
Here we present a brief outline of the method in order to discuss the 
estimator for the specific heat. For the present case, the SSE approach 
starts by casting the Hamiltonian in the form
\begin{equation}
{\hat H}=-{1\over 2}\sum_{b=1}^{3N}[{\hat H}_{1,b}-{\hat H}_{2,b}] + C,
\end{equation}
where $b$ denotes the bond connecting the nearest neighbor sites
$\langle i(b),j(b)\rangle$, $C$ is an additive constant and the operators
$H_{1,b}$ and $H_{2,b}$ are defined as 
\begin{eqnarray}
H_{1,b}&=&2J(b)[{1\over 4}-S_{i(b)}^zS_{j(b)}^z], \\
H_{2,b}&=&J(b)[S_{i(b)}^+S_{j(b)}^- + S_{i(b)}^-S_{j(b)}^+].
\end{eqnarray}
The coupling constant $J(b)=J$ for bonds in the planes and $J(b)=J_{\perp}$
for inter-planar bonds. An exact and useful expression for an operator 
expectation value at inverse temperature $\beta=J/T$,
\begin{equation}
\langle {\hat A}\rangle = {1\over Z}{\mbox{Tr}}
\lbrace {\hat A}e^{-\beta{\hat H}}\rbrace, 
\hskip 0.2in Z={\mbox{Tr}}\lbrace e^{-\beta{\hat H}}\rbrace,
\end{equation}
is obtained by expanding the density matrix $e^{-\beta {\hat H}}$ in a 
Taylor series and writing the trace as sum over the diagonal matrix elements 
in a basis 
$\lbrace |\alpha \rangle\rbrace =\lbrace |S_1^z,\ldots,S_N^z\rangle\rbrace$.
The partition function can then be written as
\begin{eqnarray}
Z &=& \sum_{n=0}^{\infty}\sum_{\alpha}\sum_{S_n}{\frac{\beta^n}{n!}}
\langle\alpha|\prod_{p=1}^n H_{a_p,b_p}|\alpha\rangle,\\
  &\equiv&\sum_{n=0}^{\infty}\beta^n\sum_{\alpha}\sum_{S_n}W'(\alpha,S_n)
\label{Zn}
\end{eqnarray}
where $S_n$ denotes a sequence of index pairs defining the operator
string $\prod_{p=1}^n H_{a_p,b_p}$:
\begin{equation}
S_n=[a_1,b_1][a_2,b_2]\dots [a_n,b_n],
\label{sn}
\end{equation}
where $a\in\{1,2,3\}$, $b\in\{1,\dots,N\}$. We have separated the temperature 
dependence of the weight factor for convenience. We can now write the 
expectation value of an operator as
\begin{equation}
\langle {\hat A}\rangle_W={1\over Z}\sum_{n=0}^{\infty}\beta^n\sum_{\alpha}
\sum_{S_n}A(\alpha,S_n)W'(\alpha,S_n).
\end{equation}
Taking ${\hat A}={\hat H}$, it can be shown\cite{sse1,handscomb} 
that the energy is given by the average length of the operator sequences
\begin{equation}
E=-{1\over \beta Z}
\sum_{n=0}^\infty n\beta^n\sum_{\alpha}\sum_{S_n}W'(\alpha,S_n)
\equiv -{1\over \beta}\langle n\rangle .
\end{equation}
A straightforward differentiation with respect to temperature gives the 
specific heat $C_v=\frac{\partial E}{\partial T}$ in the form of 
Eq.~(\ref{cnn}).

\end{document}